# The effect of electron correlation on the superconducting and normal properties of the cuprates

A. Rosencwaig*
ARIST, Danville, CA 94506

## ABSTRACT

The strong electron correlation in the cuprates can lead to an enhanced effective mass for both bosonic and fermionic quasiparticles. Where this correlation is characterized by a length that is inversely proportional to the effective temperature, the thermal wavelength for the bosonic quasiparticles becomes essentially independent of temperature. Applying this concept to a preformed pair model, such as the $(Cu)_{13}$ cluster model, gives $T_c$ vs dopant curves and superconducting gaps in good agreement with experiment. In addition, a correlation-enhanced effective mass provides a natural explanation for the anomalous normal state resistivity and optical conductivity of the cuprates.

Since the 1986 discovery of high-temperature superconductivity in the cuprates, many approaches have been tried to explain this new phenomenon but a complete understanding has remained elusive. Several models postulate preformed bosonic pairs that can undergo Bose-Einstein condensation (BEC). In one such model,[1] the preformed pairs arise from the hybridization of molecular orbitals in a cluster containing 13 Cu atoms in the $CuO_2$ plane, and the resulting bosonic quasiparticles have low binding energy and are assumed to be weakly interacting. This $(Cu)_{13}$ cluster model appears to account for many of the thermodynamic and electronic properties of the superconducting state. A key element of this model is a bosonic thermal wavelength that is independent of temperature, a condition that is difficult to envisage in a system of weakly interacting charge carriers. However, the cuprates are systems with strong electron correlations and thus one would expect that the bosonic quasiparticles would also be strongly correlated. I will show that it may be this strong correlation between the preformed pairs that results in a temperature-independent thermal wavelength.

The cuprates appear to exhibit interdependent static and dynamic charge-density-wave, CDW, and spin-density-wave, SDW, orders that give rise to static and fluctuating charge and spin stripe patterns.[2-5] These stripe patterns result from the competition between the kinetic energy of the mobile charge carriers and the superexchange interaction between neighboring $Cu^{+2}$ spins. A measure of the extent of charge and spin correlations can be obtained from the CDW-induced and the SDW-induced neutron diffraction superlattice peaks. The correlation length for a given order is the reciprocal of the linewidth of the corresponding incommensurate diffraction peak. In a study of the linewidths of the SDW peaks in $La_{2-x}Sr_xCuO_4$, LSCO, with x = 0.14, Aeppli *et al*[6] found that the linewidth, $\kappa_m$, varies as,

$$\kappa_m^2 = \kappa_{mo}^2 + \frac{T^{*2}}{a^2 T_m^2} \qquad (1)$$

and

$$T^* = \left\{ T^2 + (\hbar\omega/k)^2 \right\}^{\frac{1}{2}} \qquad (2)$$

where $T^*$ is the effective temperature, $T$ is the temperature, $\hbar\omega$ is the energy transfer, $a$ is the lattice parameter in the $CuO_2$ plane, $\kappa_{m0}$ is the linewidth (in $A^{-1}$) at zero $T^*$ and $T_m$ is an SDW correlation temperature. For LSCO with x= 0.14, Aeppli *et al* found that $\kappa_{m0} \approx .03$ $A^{-1}$ and $T_m \approx 590$ K.[6] The SDW correlation length is given by $\xi_m = 1/\kappa_m$. We can see from Eqn. (1) that for a sufficiently high $T^*$, $\xi_m$ varies inversely with the effective temperature.

Given the interdependence of the CDW and SDW orders, we would expect a similar functional relationship with respect to $T$ and $\hbar\omega$ for the CDW order, and therefore also a similar relationship for the preformed pairs in the normal state. Indeed the $(Cu)_{13}$ clusters can also order in stripes.[7] Thus the correlation length, $\xi_p$, of the $(Cu)_{13}$ clusters, or more generally, of any preformed bosonic quasiparticles will be of the form,

$$\xi_p = \frac{1}{\left\{\kappa_0^2 + \frac{T^{*2}}{a^2 T_p^2}\right\}^{1/2}} \qquad (3)$$

where $\kappa_0^{-1}$ represents the correlation length at $T^* = 0$ and $T_p$ is a pair correlation temperature, which can be defined as the temperature below which the preformed pairs begin to correlate into dynamic stripe patterns. For a sufficiently high $T^*$, the correlation length of the bosonic quasiparticles, $\xi_p$, will also be inversely proportional to the effective temperature. Note that for a study on superconductivity we can set $\hbar\omega = 0$, and then $T^* = T$.

The strong correlations will tend to increase the effective mass of the quasiparticles. Since the quasiparticles self-arrange into one-dimensional dynamical stripes, this effect should increase as the correlation length increases. In particular, if we set the effective mass of a bosonic preformed pair proportional to the correlation length, then,

$$m_b^* = m_b \frac{\xi_p}{a} \qquad (4)$$

where $m_b$ is the mass of a single non-correlated bosonic quasiparticle.

By incorporating the strong correlation between the quasiparticles into an effective mass, we can approximate the approach to Bose-Einstein condensation with the weakly interacting condition,

$$n_b \lambda_T^3 = 2.612 \qquad (5)$$

where $n_b$ is the boson density and $\lambda_T$ is the thermal wavelength which is now given by,

$$\lambda_T^2 = \frac{2\pi\hbar^2}{m_b^* kT}$$
$$= \frac{2\pi\hbar^2}{m_b kT}\left\{\eta^2 + \frac{T^2}{T_p^2}\right\}^{1/2} \qquad (6)$$

with $\eta = \kappa_0 a$, a dimensionless quantity.

We can readily see that for all temperatures $\eta T_p < T < T_p$ the thermal wavelength of the bosonic quasiparticles becomes essentially independent of $T$.

To apply this concept to the $(Cu)_{13}$ cluster model, we set the boson density, $n_b$, to,[1]

$$n_b = \frac{1}{2}\frac{n_p q}{a^2 c} P(\delta, T) \qquad (7)$$

where $q$ is the dopant charge per $Cu^{+2}$ ion in the $CuO_2$ plane, $a$ and $c$ are the lattice constants of the tetragonal primitive unit cell, $n_p$ is the number of $CuO_2$ planes per primitive unit cell, and $P(\delta, T)$ is the probability that a hole is in the bosonic ground state of the cluster with $\delta$ as the energy gap between cluster states. The probability is then,

$$P(\delta, T) = 1/\{1 + e^{-\delta/kT} + e^{-2\delta/kT} + \cdots + e^{-12\delta/kT}\} \qquad (8)$$

I have previously shown[1] that the energy gap $\delta$, which is also the binding energy of the preformed bosonic pair, is a decreasing linear function of the dopant $q$, and that $\delta \to 0$ at the dopant value $q_1$ in the overdoped region where the superconductivity disappears. Thus we can set $\delta(q) = b(q_1 - q)$, where $b$ is a pair energy parameter. Since $\delta$ is quite small, typically < 5 meV, we can set $m_b \approx 2m$ where $m$ is the electron mass. The BEC condition, Eqn.(5), then will have 3 unknowns, the pair energy parameter $b$, the dimensionless parameter $\eta$, and a pair correlation temperature $T_p$. We can evaluate these three unknowns for any given cuprate from a fit to the experimental $T_c$ vs $q$ curve. We show the results of this analysis for 5 different cuprates in Table I. The value for $\eta$ is always < $10^{-4}$. The pair correlation temperature $T_p$ is about the same for all the cuprates at $\approx 660$ K, which is $\approx \frac{1}{2}J$ where $J$ is the superexchange interaction between nearest-neighbor $Cu^{+2}$ ions.[8] From the values of $\eta$ and $T_p$, we see that the thermal wavelength of the preformed pairs is essentially independent of temperature over a very wide range, 0.1 K < $T$ < 660 K. Using the derived value of $b$ we can obtain the pairing energy $\delta^m$, and thence the superconducting gap[1] $\Delta^m = 12\delta^m$ for the optimally doped cuprates. We see from Table I that the derived superconducting gaps are in good agreement with the experimental values obtained from ARPES and tunneling spectroscopy experiments.[9]

The cuprate superconductors are distinguished not only by their high superconducting transition temperatures, but also by their anomalous transport properties in the normal state. Whereas a conventional Fermi liquid exhibits a resistivity that varies with temperature as $T^2$, the in-plane resistivity of the cuprates varies as $T$ and does so over a very wide temperature range. Similarly, the optical conductivity in the mid-IR varies with optical frequency as $1/\omega$ instead of $1/\omega^2$ as in conventional Fermi liquids. I will show that these anomalous normal properties are a natural result of the strong electron correlations in the cuprates.

Since the formation of a single preformed pair is itself a correlation between two carriers, the effective mass for an individual carrier can be written as,

$$m^* = m\frac{T_o}{T^*} \qquad (9)$$

where $T_o = 2T_p \approx 1320$ K. Thus for $T_p < T^* < T_o$, the holes first form into individual preformed pairs, such as the $(Cu)_{13}$ clusters, which have 2 holes per cluster, and then for $T^* < T_p$, the preformed pairs, or $(Cu)_{13}$ clusters, self-arrange into dynamic stripes. Throughout this process, the carrier correlation length varies as $1/T^*$.

In the context of the $(Cu)_{13}$ cluster model, the binding energy of the bosonic quasiparticles is quite low, typically < 5 meV,[1] and thus for temperatures above the superconducting transition temperature, $T_c$, the clusters are primarily fermionic quasiparticles, since $\delta < kT_c$. Therefore, we can treat the normal state as being essentially a Fermi liquid with strong electron correlations characterized by a correlation length that is proportional to $1/T^*$.

The dc conductivity, $\sigma(0)$, for this Fermi liquid is then,

$$\sigma(0) = \frac{ne^2\tau(T)}{m^*} \qquad (10)$$

where $n$ is the carrier density and $\tau$ the relaxation time. For a Fermi liquid, $\tau \propto 1/T^2$. But from Eqn. (9), $m^* \propto 1/T$ when $\omega = 0$. Therefore, $\sigma(0) \propto 1/T$, and the resistivity $\rho(T) \propto T$ up to the temperature $T_o$ or $\approx 1320$ K.

The real part of the optical conductivity, $\sigma(\omega)$, in the mid-IR is given by

$$\sigma(\omega) = \frac{ne^2\tau(\omega)}{m^*} \qquad (11)$$

In the mid-IR, $\hbar\omega > kT$, and so from Eqn. (9), $m^* \propto 1/\omega$. Since $\tau(\omega)$ for a Fermi liquid $\propto 1/\omega^2$, we can see from Eqn. (11) that $\sigma(\omega) \propto 1/\omega$. Defining the effective relaxation rate, $\dfrac{1}{\tau^*(\omega)} = \dfrac{m^*}{m}\dfrac{1}{\tau(\omega)}$, we then have the well-known experimental result for the cuprates, $1/\tau^*(\omega) \propto \omega$. More generally, the effective relaxation rate will vary linearly with the effective temperature, a result also in agreement with experiment, although the effective temperature is now given by $T^* = T + \hbar\omega/k$ rather than by Eqn. (2).[10]

In conclusion, I have shown that the strong electron correlation in the cuprates can lead to an enhanced effective mass for both the bosonic and fermionic quasiparticles. Where this correlation is characterized by a length that is inversely proportional to the effective temperature, the thermal wavelength for the bosonic quasiparticles becomes essentially independent of temperature. Applying this concept to a preformed pair model, such as the (Cu)$_{13}$ cluster model, gives $T_c$ vs $q$ curves and superconducting gaps that are in good agreement with experiment. In addition, the principal characteristics of both the resistivity and the optical conductivity of the normal state of the cuprates are readily explained by the concept of a Fermi liquid with correlation-enhanced carrier effective mass. Although specifically applied here to the cuprates, this correlation-enhanced effective mass concept might also apply to other Bose-Einstein condensation systems where there are strong interactions or correlations between the bosonic particles.

## ACKNOWLEDGMENTS

I would like to acknowledge helpful discussions with M. Greven and A. Jacobs.

TABLE I: The best-fit values for $\eta$, $T_0$ and $b$ from the experimental $T_c$ vs $q$ curves for several cuprates; and the calculated pairing energy, $\delta^m$, the calculated superconducting gap $\Delta^m = 12\delta^m$, and the experimental values of $\Delta^m$ (Ref. (9)) for the optimally doped cuprates.

| **Cuprate** | $\eta$ | $T_0$ (K) | $b$ (meV) | $\delta^m$ (meV) | $\Delta^m$ (meV) | $\Delta^m_{exp}$ (meV) |
|---|---|---|---|---|---|---|
| La$_{2-x}$Sr$_x$CuO$_4$ (LSCO) | $< 10^{-4}$ | 670 | 14.8 | 1.63 | 19.6 | 20 |
| Nd$_{2-x}$Ce$_x$CuO$_4$ (NCCO) | $< 10^{-4}$ | 660 | 25.9 | 1.03 | 12.4 | 12 |
| YBa$_2$Cu$_3$O$_{7-y}$ (Y123) | $< 10^{-4}$ | 665 | 28.7 | 3.44 | 41.3 | 40 |
| Bi$_2$Sr$_2$CaCu$_2$O$_{8-y}$ (Bi2212) | $< 10^{-4}$ | 590 | 33.8 | 4.03 | 48.4 | 42 |
| Hg$_1$Sr$_2$Ba$_2$Cu$_3$O$_{10-y}$ (Hg1223) | $< 10^{-4}$ | 670 | 40.5 | 4.86 | 58.3 | 54 |